\documentclass[aps,a4paper,twocolumn,nofootinbib,superscriptaddress]{revtex4}

\usepackage[english]{babel}

\usepackage[utf8]{inputenc}

\usepackage{amsmath,amssymb,amsfonts,mathrsfs}

\usepackage[amsmath,thmmarks]{ntheorem}

\usepackage{graphicx}
\usepackage{siunitx}
\DeclareSIUnit\Molar{\textsc{M}}

\usepackage[linkcolor=black,colorlinks=true,citecolor=black,filecolor=black]{hyperref}
\usepackage{dsfont}

\usepackage{physics}

\def\ec/{\textit{E.coli}}
\newcommand{\Drot}{D_{\mathrm{rot}}}
\newcommand{\lequ}{\lambda_{\mathrm{equ}}}

\renewcommand{\vec}[1]{\mathbf{#1}}
\newcommand{\mymatrix}[1]{{\mathit{#1}}}

\begin{document}

\title{Traveling concentration pulses of bacteria in a generalized Keller-Segel model}



\author{Maximilian Seyrich}
\email[Email: ]{seyrich@tu-berlin.de}
\affiliation{Institut f\"ur Theoretische Physik, Technische Universit\"at Berlin, Hardenbergstrasse 36, 10623 Berlin, Germany} 

\author{Andrzej Palugniok}
\email[Email: ]{andrzej.palugniok@worc.ox.ac.uk}
\affiliation{Worcester College, University of Oxford, Walton Street, OX1 2HB Oxford, United Kingdom}
\affiliation{Institut f\"ur Theoretische Physik, Technische Universit\"at Berlin, Hardenbergstrasse 36, 10623 Berlin, Germany}

\author{Holger Stark}
\email[Email: ]{holger.stark@tu-berlin.de}
\affiliation{Institut f\"ur Theoretische Physik, Technische Universit\"at Berlin, Hardenbergstrasse 36, 10623 Berlin, Germany}

\date{\today} 

\begin{abstract}
We formulate the Smoluchowski equation for a run-and-tumble particle. It includes the mean tumble rate in a chemical field, for which we derive a Markovian response theory. Using a multipole expansion and a reaction-diffusion equation for the chemoattractant field, we derive a polarization extended model, which also includes the recently discovered angle bias. In the adiabatic limit we recover generalized Keller-Segel equations with diffusion and chemotactic coefficients that depend on the microscopic swimming parameters. By requiring the tumble rate to be positive, our model possesses an upper bound of the chemotactic drift velocity, which is no longer singular as in the original Keller-Segel equations. Using the Keller-Segel model, we present an extensive study of traveling bacterial concentration pulses demonstrating how speed, width, and height of the pulse depend on the microscopic parameters. Most importantly, we discover a maximum number of bacteria that the pulse can sustain - the maximum carrying capacity. Finally, we obtain a remarkably good match to experimental results on the traveling bacterial pulse. It does not require a second, signaling chemical field nor a singular chemotactic drift velocity.
\end{abstract}

\maketitle


\section{Introduction}


Collective motion of biological and artificial microswimmers is a most appealing research field as demonstrated in several review articles \cite{ramaswamy2010mechanics, romanczuk2012active,marchetti2013hydrodynamics,zottl2016emergent,bechinger2016active}. The formation of various patterns and clustering have been investigated both experimentally and theoretically
in systems of bacteria \cite{adler1966chemotaxis, budrene1991complex, mittal2003motility,sokolov2007concentration, tailleur2008statistical,zhang2010collective,dombrowski2004self,reinken2018derivation}, of eukaryotic cells such as \textit{Dictyostelium discoideum} or human sperm \cite{rothschild1948activity,bonner1947evidence, palsson2000model,ben2000cooperative,strassmann2000altruism, kaiser2003coupling, riedel2005self,friedl2009collective, schoeller2018flagellar}, as well as in suspensions of active colloids \cite{saintillan2008instabilities,theurkauff2012dynamic, buttinoni2013dynamical, zottl2014hydrodynamics,pohl2014dynamic, speck2014effective, liebchen2015clustering, pohl2015self,zottl2016emergent, blaschke2016phase,kuhr2017collective}.
In this article we study the collective behavior of a bacterial population, which in the concentration field of a chemoattractant forms a traveling solitary pulse.

The motility mechanism of the run-and-tumble bacterium \ec/ has been extensively studied \cite{berg1972chemotaxis,block1982impulse,berg2008coli,saragosti2011directional,sourjik2012responding,masson2012noninvasive, pohl2017inferring,seyrich2018statistical}. 
Bacteria perform chemotaxis, the ability to sense and respond to chemical gradients in order to find better living conditions. They realize the chemotactic drift motion along a chemical gradient by elongated run phases if the environment becomes more favorable while runs are shortened in the opposite case. 
The internal chemotaxis machinery of the bacterium senses and compares the nutrient concentration in time, which is rationalized in a linear response theory for the tumble rate \cite{macnab1972gradient, block1982impulse, de2004chemotaxis, eisenbach2004chemotaxis}. More recently, a second chemotaxis strategy, called angle bias, has been reported \cite{saragosti2011directional,pohl2017inferring,seyrich2018statistical}.
The mean reorientation angle during tumbling is reduced if the bacterium swims along a chemical gradient and increased in the opposite case. This also generates a net drift motion in the favorable direction. Finally, using logarithmic sensing, \ec/ is able to perform chemotaxis in concentration fields varying by many orders of magnitude \cite{dahlquist1972quantitative,KALININ20092439,lazova2011response}.
Such an ability is commonly described by Weber's law in different physical areas \cite{weber1834pulsu,fechner2012elemente}.

A very interesting collective phenomenon in a bacterial population is a concentration pulse that travels along a capillary tube with almost no dispersion nearly like a soliton \cite{adler1966chemotaxis, saragosti2011directional}, 
most recently also observed in a population with non-genetic variations \cite{fu2018spatial}.
The pulse is initiated in an initially uniform environment of a chemoattractant. A bacterial population concentrated in space eats the nutrient and thereby creates a chemical gradient along which it drifts towards untouched regions. Moreover, Adler in his experiments also observed that not all bacteria travel with the pulse but are left behind at the initial location \cite{adler1966chemotaxis}, which indicates a finite carrying capacity of the traveling pulse. Further chemoattractants present in Adler's experiments then initiated further pulses emerging from the bacteria left behind.

A very prominent theoretical approach to describe the traveling bacterial pulse is the celebrated Keller-Segel model \cite{keller1971traveling}, originally introduced for the aggregation of slime molds \cite{keller1970initiation}. It couples a diffusion-drift equation for the bacterial density to a reaction equation for the nutrient. However, the Keller-Segel model has two drawbacks. First, a soliton solution (classified as unstable \cite{nagai1991traveling}) only occurs if the chemotactic drift velocity diverges for vanishing nutrient concentration. Second, nutrient diffusion was neglected. Later, based on analytic arguments, Ref. \cite{nagai1991traveling} demonstrated that traveling pulses 
also exist in the presence of nutrient diffusion. More importantly, Brenner \emph{et al.} showed that the singularity in the chemotactic drift velocity is not necessary if one introduces a second chemoattractant, which the bacteria excrete themselves \cite{brenner1998physical}. Reference \cite{saragosti2011directional} followed this approach to formulate a kinetic model (inspired by Ref.\ \cite{othmer1988models}), which describes traveling pulses in their experiments. Finally, a modification of this kinetic model has recently been used to investigate pulse propagation in the presence of two \ec/ populations \cite{emako2016traveling}. The Keller-Segel equations find wide applications in modeling bacterial chemotaxis as reviewed in Ref.\ \cite{tindall2008overview}. They have also been derived for active Brownian particles, which propel by self-diffusiophoresis, and for quorum-sensing run-and-tumble particles \cite{rein2016collective}.

Multipole expansions have frequently been applied to microswimmers in order to approximate the Smoluchowski equation for the full distribution function in the microswimmer's position and orientation \cite{golestanian2012collective,pohl2014dynamic,stark2016swimming, rein2016collective,reinken2018derivation}.
Besides for density such expansions also provide an additional dynamic equation for the polarization, which unraveled interesting collective behavior of Janus particles \cite{liebchen2015clustering} and which also allowed to investigate steady-state distributions of run-and tumble particles \cite{schnitzer1993theory}. 
Our derivation is inspired by the approach of the latter reference but extends it by introducing the concentration field of a chemoattractant.

In this article we formulate a Markovian response theory for the tumble rate. It includes logarithmic sensing for which we introduce an upper threshold. We use the tumble rate in the Smoluchowski equation and derive a polarization extended 
model (PE) to treat chemotaxis of non-interacting \ec/ bacteria. The PE model
contains equations for the bacterial density, the bacterial polarization, and the chemical concentration field. 
In a second step, we also include the recently discovered angle bias. In the adiabatic limit the PE model simplifies to a generalized Keller-Segel model (KS) 
where the coefficients for diffusion and chemotactic drift velocity depend on the microscopic swimming parameters
of the bacterium. In particular, the chemotactic coefficient is not singular in the chemical concentration. We numerically solve both models for an initially uniform chemoattractant and a bacterial population concentrated in space using parameters that are realistic for the \ec/ bacterium. The traveling bacterial pulse generated by both the PE and KS model are identical thus the KS model is a valid approximation of the full kinetic formalism. We present a detailed parameter study of the traveling pulse and identify a maximum carrying capacity as a consequence of the bounded chemotactic drift velocity, which has not been mentioned so far. It means that the pulse can only sustain a finite number of bacteria. Finally, we tune our parameters to match the experimental realization of the bacterial pulse in Ref. \cite{saragosti2011directional}. Hence, our generalized Keller-Segel model is able to describe traveling bacterial pulses without the need neither for a singular chemotactic drift velocity nor for a second chemoattractant.

The remainder of the article is organized as follows. We present the Markovian response theory for the tumble rate in Sect. \ref{sec:MeanFieldTumbleRate}. 
We use it to derive the polarization extended model (PE) and the generalized Keller-Segel model (KS) in 
Sects. \ref{sec:PEModel} and \ref{sec:KS}.
We also incorporate the angle bias and formulate a non-dimensional version of the KS model in Sects.\ \ref{sec:AngleBias} and \ref{sec:rescaling}. Details of the numerical solution scheme are given in Sect. \ref{sec:NumSim} and Sect. \ref{sec:ChapterIV} presents our detailed numerical study. We close with conclusions and an outlook in Sect.\ \ref{sec:conclusion}.


\begin{figure}
\includegraphics[width=0.5\textwidth]{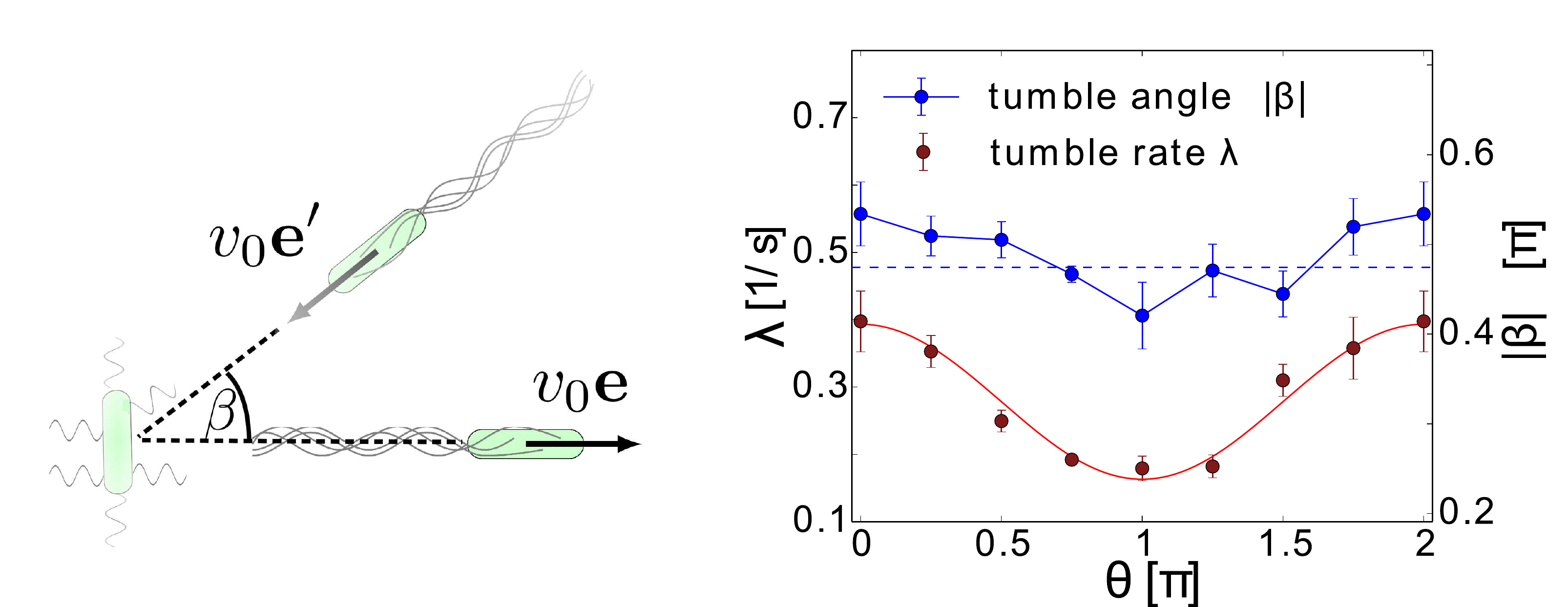}
\caption{Left: Schematic of an individual \ec/ tumbling from its previous direction $\vec{e'}$ to $\vec{e}$ with tumble angle $\beta$. Right: Mean tumble rate $\lambda(\theta)$ (red) as a function of the swimming angle $\theta$
measured against the negative chemical gradient. The tumble rate was averaged over a population of around 1.000 \ec/ in a linear gradient of $\alpha$-methyl-aspartate. The fit with a cosine function, $y(x)= a_1 + a_2 \cos(x)$ (red line) according to Eq.\ (\ref{eq:tumbleratefinal}), yields regression parameters $a_1=0.278$ and $a_2=0.115$.
The blue line represents the mean tumble angle $\langle\cos{\beta}\rangle(\theta)$ as a function of $\theta$ of the same 
population. It indicates an angle bias for tumbling in a chemical gradient. 
Adapted from Ref.\ \cite{pohl2017inferring}.}
\label{fig:fit1}
\end{figure}

\section{Model}
\label{sec:ChapterII}

\subsection{Markovian response theory for tumble rate}
\label{sec:MeanFieldTumbleRate}
Bacteria tumble less when moving up a chemical gradient. Based on the established linear-response theory, we formulate an equation for the tumble rate $\lambda(\vec{r},\vec{e})$ as a function of the swimming direction $\vec{e}$. Below, we will relate it to the angle $\theta$ relative to the local gradient $\grad c$ of a chemoattractant with density $c$.

We start with the linear-response theory. It gives the tumble rate $\lambda(t)$ as a function of time and depends on the bacterium's past trajectory $\vec{r}(t')$,
\begin{equation}
\label{eq:tumblerate}
\lambda(t) = \lequ  - \int\limits_{-\infty}^{t}  R(t-t') \, c(\vec{r}(t')) dt'~,
\end{equation}
where we have introduced the response kernel $R(t)$ and $\lequ$ is the tumble rate without any chemical gradient.
Note that Eq.\ (\ref{eq:tumblerate}) describes a non-Markovian process. We convert it to a Markovian process with $\lambda$ depending on location $\vec{r}$ and swimming direction $\vec{e}$ by averaging over all possible histories, which gives the mean history for a given swimming direction $\vec{e}$. To do so, we split the integral on the R.H.S. of Eq.\ (\ref{eq:tumblerate}) into contributions from individual runs, during which the according swimming directions $\vec{e}_i$ are assumed to be constant. Averaging over the history of all possible paths, we can show that each of these contributions gives a term proportional to the scalar product $\vec{e}\cdot\frac{\grad{c}}{c}$.
We also used here that the response kernel is proportional to the inverse background conversation, $R \propto 1/c $, which was indeed measured in experiments \cite{masson2012noninvasive}. For the detailed derivation we refer to appendix \ref{sec:MarkovianTumblerate} and present the final result,
\begin{equation}
\label{eq:tumbleratefinal}
\lambda(\vec{r},\vec{e}) = \lequ - \chi_0v_0\vec{e}\cdot\frac{\grad{c(\vec{r})}}{c(\vec{r})} \, .
\end{equation}
Here, $v_0$ is the swimming velocity of the bacterium and $\chi_0$ is a unitless measure of the chemotactic strength. It depends on integrals over the response function $R$ and moments of the tumble angle distribution $P(\beta)$. Note that we obtain here $\lambda \propto \grad (\ln c)$ commonly known as logarithmic sensing and Weber's law. 
It was measured, for example, in Ref. \cite{lazova2011response}.

When we define the orientation angle relative to the negative chemical gradient, $\cos\theta = - \vec{e}\cdot\frac{\grad{c}}{|\grad{c}|}$, the tumble rate becomes
\begin{equation}
\label{eq:tumbleratefinal2}
\lambda(\vec{r}, \theta)
=  \lequ + \chi_0 v_0 \frac{|\grad{c(\vec{r})}|}{c(\vec{r})} \cos{\theta}  \, .
\end{equation}
In Fig.\ \ref{fig:fit1} adapted from Ref.\ \cite{pohl2017inferring}, the red points show experimental data for the mean tumble rate $\lambda(\vec{r},\theta)$. It was obtained by averaging over a population of around $1000$ individual bacteria in a linear gradient. The appropriate cosine fit (red line) confirms our theoretically derived result of Eq.\ (\ref{eq:tumbleratefinal}).

In general, Eq.\ (\ref{eq:tumbleratefinal}) can produce a negative tumble rate for a sufficiently large gradient of $\log c$, which is even singular at $c=0$. The reason is the relation $R \propto 1/c$ for the response function mentioned earlier. While this dependence was measured for a wide range of background concentrations
\cite{masson2012noninvasive}, clearly the deviation from $\lequ$ in Eq. (\ref{eq:tumbleratefinal}) has to saturate to a value smaller than $\lequ$.
Furthermore, it is known that the bacterium needs a small threshold concentration $c_t$ to perform
chemotaxis \cite{adler1969chemoreceptors}. To implement both constraints we use the hyperbolic tangent function in Eq. (\ref{eq:tumbleratefinal}) and write,
\begin{equation}\label{eq:tumbleratebounded}
    \lambda(\vec{r},\vec{e}) = \lequ - \chi_0\frac{v_0}{\delta} 
   \tanh{\left(\frac{c}{c_t} \right)}
    \tanh{\left(\delta\frac{|\grad{c}|}{c}\right)}\vec{e}\cdot\hat{\vec{s}}  \, .
\end{equation}

Here, we have introduced $\hat{\vec{s}} = \frac{\grad{c}}{|\grad{c}|}$ and do not explicitly state the space dependence of the concentration $c$. This expression recovers Eq.\ (\ref{eq:tumbleratefinal}) for $\delta |\grad{c}| / c < 1$ and $c>c_t$, while it smoothly approaches 
the minimum tumble rate $\lequ - \chi_0\tfrac{v_0}{\delta}$ at a large relative chemical gradient or it becomes $\lequ$ for $c < c_t$. The chemotactic length $\delta$ quantifies the strength of the logarithmic derivative of $c(\vec r)$. In the following, we use the notation 
\begin{equation}
\chi\left(\frac{|\grad{c}|}{c}\right) :=\chi_0\frac{v_0}{\delta} \tanh{\left( \frac{c}{c_t}\right)} \tanh{\left(\delta\frac{|\grad{c}|}{c}\right)} \, .
\end{equation}

\subsection{Polarization extended model (PE)}
\label{sec:PEModel}

\textit{Smoluchowski equation.} -- We first construct dynamic equations for the evolution of the one-particle distribution function $\psi(\vec{r},\vec{e},t)$ of position $\vec{r}$ and orientation $\vec{e}$ at time $t$ and the concentration of chemoattractant, $c(\vec{r},t)$. We begin with a generalized Smoluchowski equation for $\psi$, which contains the usual contributions from translational and rotational currents, $\vec{J}_{\mathrm{trans}}$ and $\vec{J}_{\mathrm{rot}}$, but also contributions from tumble events represented by $\mathcal{F}\{\psi\}$ and from cell division and death, $\frac{r}{S_d}\int\psi(\vec{r},\vec{e}',t) d\vec{e}'$:
\begin{align} 
\begin{aligned}
    \frac{\partial\psi}{\partial t} =& -\div{\vec{J}_{\mathrm{trans}}} 
    + -\mathcal{R}\cdot\vec{J}_{\mathrm{rot}}\\
    &+ \mathcal{F}\{\psi\} + \frac{r}{S_d}\int\psi(\vec{r},\vec{e}',t) d\vec{e}' \, .
\end{aligned}
\end{align}
Here, $\mathcal{R}=\vec{e}\times\partial_{\vec{e}}$ and $r = \ln{(2)}/\tau$ is the net creation rate with $\tau$ being the mean doubling time of bacterial cells \cite{saragosti2011directional}. Here, we assume that the net creation of cells does not depend of their directon $\vec e$ and $S_d$ is the surface area of a $d$ dimensional unit sphere (full solid angle). For the translational current we include active motion and translational diffusion, $\vec{J}_{\mathrm{trans}}=v_0\vec{e}\psi-D\grad{\psi}$, where $D$ is the translational diffusion coefficient and $v_0$ is the bacterial swimming speed. The rotational current is purely diffusive, $\vec{J}_{\mathrm{rot}}=-\Drot\mathcal{R}\psi$, where $\Drot$ is the rotational diffusion coefficient. According to Ref.\ \cite{berg1972chemotaxis} we take a Poisson distribution for the run times and so write the term for the tumble events as
\begin{align} \label{eq:tumble-events}
\begin{aligned}
\mathcal{F}\{\psi\}=&-\lambda(\vec{r},\vec{e})\psi \\
&+ \int P(\vec{e}-\vec{e}',\vec{e}')\lambda(\vec{r},\vec{e}')\psi(\vec{r},\vec{e}',t)d\vec{e}' \, .
\end{aligned}
\end{align}
We introduced the tumble rate $\lambda(\vec{r},\vec{e})$ and $P(\vec{e}-\vec{e}',\vec{e}')$ is the probability of a bacterium to reorient from orientation $\vec{e}'$ to $\vec{e}$.
In Eq.\ (\ref{eq:tumble-events}) the first term on the R.H.S. represents events, which cause bacteria with orientation $\vec{e}$ to tumble into any orientation, and the second term represents all events, which cause bacteria with other orientations to tumble into orientation $\vec{e}$. The complete Smoluchowski equation for the evolution of $\psi$ now reads
\begin{align} 
	\begin{aligned} \label{eq:master-dens}
			\frac{\partial\psi}{\partial t} = \ &-\div{ (v_0\vec{e}\psi)} \ + D\nabla^2\psi  + \Drot\mathcal{R}^2\psi\\
			&- \lambda(\vec{r},\vec{e})\psi + \int P(\vec{e}-\vec{e}',\vec{e}')\lambda(\vec{r},\vec{e}')\psi(\vec{r},\vec{e}',t)d\vec{e}'\\ &+\frac{r}{S_d}\int\psi(\vec{r},\vec{e}',t) d\vec{e}' \,,
	\end{aligned}
\end{align}

For completeness, we also write a reaction-diffusion equation for the chemoattractant concentration $c$, which is also consumed by bacteria with constant rate $k$,
\begin{align} 
	&\begin{aligned} \label{eq:master-conc}
		\frac{\partial c}{\partial t} = D_c \nabla^2c - k
		\int\psi(\vec{r},\vec{e}',t)d\vec{e}'.
	\end{aligned}
\end{align}\\


\textit{Multipole expansion.} -- In order to make progress with Eqs. (\ref{eq:master-dens}) and (\ref{eq:master-conc}), 
we assume that the probability distribution for a specific tumble event does not depend on the initial orientation of the bacterium, $P(\vec{e}-\vec{e}',\vec{e}')=P(\vec{e}-\vec{e}')$. Therefore, we can write for the zeroth and first moment,
\begin{gather}
	\begin{aligned} \label{eq:distnorm}
		\int P(\vec{e}-\vec{e}')d\vec{e} = 1,
	\end{aligned}\\
	\begin{aligned} \label{eq:distavg}
		\int \vec{e}P(\vec{e}-\vec{e}')d\vec{e} = \langle\cos{\beta}\rangle\vec{e}',
    \end{aligned}
\end{gather}
where $\langle\cos{\beta}\rangle$ is the mean of the cosine of the reorientation angle $\beta$ \cite{schnitzer1993theory}. We take the tumble rate $\lambda(\vec{r},\vec{e})$ to vary as in Eq.\ (\ref{eq:tumbleratebounded}). Now, we integrate Eq. (\ref{eq:master-dens}) over all orientations $\vec{e}$ and define the bacterial density $\rho(\vec{r},t)=\int\psi(\vec{r},\vec{e},t)d\vec{e}$ and polarization $\vec{P}(\vec{r},t)=\int\vec{e}\psi(\vec{r},\vec{e},t)d\vec{e}$, which correspond to the zeroth and first moments of $\psi$ respectively, to obtain
\begin{align} \label{eq:dens}
	\frac{\partial\rho}{\partial t} = - \div{ (v_0\vec{P})} \  + D\nabla^2\rho +r\rho \, .
\end{align}
We have also used $\int\mathcal{R}^2\psi d\vec{e}=0$ and the normalization condition Eq.\ (\ref{eq:distnorm}) to show that tumbling does not contribute to Eq.\ (\ref{eq:dens}), as it should be.

In order to derive a dynamic equation for the polarization, we compute \mbox{$\int \vec{e} \,$Eq.\ (\ref{eq:master-dens})$\,d\vec{e}$} and introduce the quadrupole moment $\mymatrix{\mathbf{Q}} = \int(\vec{e}\otimes\vec{e} - \frac{\mathbf{1}}{d})\psi(\vec{r},\vec{e},t)d\vec{e}$, with $d$ being the number of spatial dimensions. This gives
\begin{align} \label{eq:polQ}
\begin{aligned}
		\frac{\partial\vec{P}}{\partial t} = &-v_0
		\mathbf{\boldsymbol{\nabla} Q}
		- \frac{v_0}{d}\grad{\rho} + D\nabla^2\vec{P} \\ 
		&-[\Drot(d-1)+(1-\langle\cos{\beta}\rangle)\lequ]\vec{P} \\ 
		&+(1-\langle\cos{\beta}\rangle)\chi\left(\frac{|\grad{c}|}{c}\right)\mymatrix{\mathbf{Q}}\,
		\hat{\vec{s}}\\
		&+ \frac{1-\langle\cos{\beta}\rangle}{d}\chi\left(\frac{|\grad{c}|}{c}\right)\rho\ \hat{\vec{s}},
\end{aligned}
\end{align}
where we used $\mathcal{R}^2\vec{e}=-(d-1)\vec{e}$ and Eq.\ (\ref{eq:distavg}). To truncate the multipole expansion, we neglect the 
qua\-dru\-pole moment $\mymatrix{\mathbf{Q}}$ and define the relaxation rate
\begin{equation}
\label{eq.omega}
\omega=\Drot(d-1)+(1-\langle\cos{\beta}\rangle)\lequ \, ,
\end{equation}
on which polar order relaxes or decorrelates in time. Thus, we ultimately obtain
\begin{align} \label{eq:pol}
\begin{aligned}
		\frac{\partial\vec{P}}{\partial t} = 
		& - \omega\vec{P} + D\nabla^2\vec{P} - \frac{v_0}{d}\grad{\rho} \\ 
	& + \frac{1-\langle\cos{\beta}\rangle}{d}\chi\left(\frac{|\grad{c}|}{c}\right)\rho\ \hat{\vec{s}} \, .
\end{aligned}
\end{align}

Finally, with our definition of bacterial density $\rho$, we can write Eq.\ (\ref{eq:master-conc}) in a simpler form,
\begin{align} \label{eq:conc}
		\frac{\partial c}{\partial t} = D_c \nabla^2c - k\rho \, .
\end{align}

\subsection{The Keller-Segel model as adiabatic limit}
\label{sec:KS}

In the case of high Peclet numbers ($\mathrm{Pe} = a v_0 / D \gg 1$), where we can neglect translational diffusion, and on large times $t \gg \frac{1}{\omega}$, where the adiabatic limit $\tfrac{\partial\vec{P}}{\partial t}\approx 0$ applies, the polarization from Eq.\ (\ref{eq:pol}) becomes
\begin{align} \label{eq:pol-closed}
    \vec{P}= - \frac{v_0}{\omega d}\grad{\rho} 
		 + \frac{1-\langle\cos{\beta}\rangle}{\omega d} \chi\left(\frac{|\grad{c}|}{c}\right)\rho\ \hat{\vec{s}} \,.
\end{align}
We remind that
$\hat{\vec{s}} = \grad{c} / |\grad{c}|$.
Substituting Eq.\ (\ref{eq:pol-closed}) into Eq.\ (\ref{eq:dens}), we obtain the generalized Keller-Segel model
\begin{align} 
	&\begin{aligned} \label{eq:dens-KS}
		\frac{\partial \rho}{\partial t} =&\ D_{\mathrm{eff}}\nabla^2\rho +r\rho\\
		&-\frac{v_0(1-\langle\cos{\beta}\rangle)}{\omega d} \div{\left[\chi\left(\frac{|\grad{c}|}{c}\right)\rho\ \hat{\vec{s}}\right]},
	\end{aligned}\\[3pt]
\vspace{0.1cm}
	&\begin{aligned} \label{eq:conc-KS}
		\frac{\partial c}{\partial t} = D_c \nabla^2c - k\rho,
	\end{aligned}
\end{align}
where $D_{\mathrm{eff}}=v_0^2/\omega d + D$ is the typical translational diffusion coefficient of an active particle,
the orientation of which decorrelates on the characteristic time $\omega^{-1}$.

From the third term on the R.H.S. of Eq. (\ref{eq:dens-KS}) we read off the chemotactic velocity along the chemical gradient,
\begin{equation}
\label{eq:ChemDrift}
\vec{v}_{\text{ch}} =
\frac{v_0(1-\langle\cos{\beta}\rangle)}{\omega d} 
\chi\left(\frac{|\grad{c}|}{c}\right) \hat{\vec{s}} \, .
\end{equation}
Taking $\chi = \chi_0 v_0 |\grad{c}|/c$, we recover the model suggested by Keller and Segel with $\vec{v}_{\text{ch}} \propto \grad{c} / c$ \cite{keller1971traveling}, and the chemotactic drift velocity is determined by a combination of
microscopic parameters, $\tfrac{\chi_0v_0^2(1-\langle \cos\beta\rangle)}{\omega d}$, called the chemotactic constant.
However, as stated earlier, according to Eq.\ (\ref{eq:tumbleratefinal}) the form for $\chi$ implies negative tumble rates for small $c$ and sufficiently steep chemoattractant gradient. The maximum value that $\chi$ can physically assume is $\lequ$, where the tumble rate becomes negative. As a result, the chemotactic speed $v_{\text{ch}} = |\vec{v_{\textrm{ch}}}|$ is also bounded. Taking $\chi = \lequ$ and approximating $\omega\approx(1-\langle\cos{\beta}\rangle)\lequ$ since $\Drot$ in Eq.\ (\ref{eq.omega}) is usually much smaller than $\lequ$, we find
\begin{align}
\label{eq:BoundedDrift}
v_{\text{ch}} \le \frac{v_0}{d} \, . 
\end{align}
This shows that an appropriately bounded tumble rate is closely linked to a physically bounded chemotactic drift.


\subsection{Bias of tumble angles}
\label{sec:AngleBias}

Up to this point we have ignored the effect of a bias in the tumble angle towards smaller mean values when swimming up the chemical gradient. This has recently been observed in experiments \cite{saragosti2011directional,pohl2017inferring}.
We now introduce it by allowing the probability distribution for a specific tumble event to explicitly depend on the initial orientation of the bacterium,
$\vec{e'}$. Hence Eqs. (\ref{eq:distnorm}) and (\ref{eq:distavg}) become
\begin{gather}
	\begin{aligned} \label{eq:distnorm-ext}
		\int P(\vec{e}-\vec{e}',\vec{e}')d\vec{e} = 1,
	\end{aligned}\\
	\begin{aligned} \label{eq:distavg-ext}
		\int \vec{e}P(\vec{e}-\vec{e}',\vec{e}')d\vec{e} = \langle\cos{\beta}\rangle(\vec{e}' ) \, \vec{e}' \, .
	\end{aligned}
\end{gather}
Equation\ (\ref{eq:distnorm-ext}) states, whatever the initial orientation of the bacterium the respective distribution is normalized. In Eq.\ (\ref{eq:distavg-ext}) the value of the mean cosine of the tumble angle $\langle\cos{\beta}\rangle(\vec{e}')$ now explicitly depends on the initial orientation $\vec{e}'$ before the tumble event.

The effect of an angle bias is to lower the mean tumble angle when the bacterium aligns with the chemoattractant gradient, hence the value of $\langle\cos{\beta}\rangle(\vec{e}')$ will increase for stronger alignment. 
Expanding $\langle\cos{\beta}\rangle(\vec{e}')$ up to the first Legendre polynomial, thus taking into account the leading polar correction, yields 
\begin{gather} 
	\begin{aligned} \label{eq:distavg-persist}
          \langle\cos{\beta}\rangle(\vec{e}') = \langle\cos{\beta}\rangle_0 + \sigma  \left(\frac{|\grad{c}|}{c}\right)          
          \vec{e}'\cdot\hat{\vec{s}} \, ,
	\end{aligned}
\end{gather}
where $\sigma$ is a positive and monotonically increasing function. It is bounded such that its maximum value 
$\sigma_{\mathrm{max}} \leq 1-\langle\cos{\beta}\rangle_0$, with $\langle\cos{\beta}\rangle_0$ being the mean cosine of the tumble angle, when the angle bias is not taken into account. Using Eqs.\ (\ref{eq:tumbleratefinal}), (\ref{eq:distnorm-ext}), (\ref{eq:distavg-ext}), and (\ref{eq:distavg-persist}), we can retrace the steps of the multipole expansion (see appendix\ \ref{subsec.app.bias} for details) to obtain an extended form for Eq. (\ref{eq:pol}) with Eqs. (\ref{eq:dens}) and (\ref{eq:conc}) remaining unchanged,
\begin{align}
	\begin{aligned} \label{eq:pol-ext}
		\frac{\partial\vec{P}}{\partial t} = &- \bigg\{ \omega +  \left[\frac{1}{d+2} \chi \left(\frac{|\grad{c}|}{c} \right)
		               \sigma \left(\frac{|\grad{c}|}{c}\right) \right]  \\		
                 & \qquad \times (\mathbf{1} + 2   \hat{\vec{s}} \otimes  \hat{\vec{s}} ) \bigg\} \vec{P} 
                     + D\nabla^2\vec{P} - \frac{v_0}{d}\grad{\rho} \\ 
		& + \bigg[\frac{1-\langle\cos{\beta}\rangle_0}{d}\chi\left(\frac{|\grad{c}|}{c}\right)
		   + \frac{\lequ}{d}\sigma\left(\frac{|\grad{c}|}{c}\right) \bigg] \rho\ \hat{\vec{s}} \, .\\
 	\end{aligned}
\end{align}
One immediately recognizes that the angle bias renormalizes the relaxation rate of the polarization and makes it anisotropic.
Thus polarizations along and perpendicular to the chemical gradient relax with different rates. In the adiabatic limit 
$\frac{\partial \vec{P}}{\partial t} \approx 0$ and for large $\mathrm{Pe}$ we can again solve for the polarization by inverting the matrix in front of $\vec{P}$ in Eq.\ (\ref{eq:pol-ext}). Substituting the resulting equation into Eq.\ (\ref{eq:dens}), we again obtain a generalized Keller-Segel equation and a chemotactic velocity $\vec{v}_{\text{ch}}$ along the chemical gradient. It now also depends on the angle bias quantified by $\sigma$. We refrain from giving the lengthy expression.


\subsection{Rescaling the Keller-Segel equations}
\label{sec:rescaling}
In order to identify essential parameters especially in the generalized Keller-Segel equations (\ref{eq:dens-KS}) and 
(\ref{eq:conc-KS}), we introduce unitless quantities.  
First, we rescale time with the chemical consumption rate, $\tilde{t} = k t$, 
lengths by the distance $l = (\tfrac{v_0^2}{\omega d k})^{1/2}$, by which a bacterium diffuses in time $k^{-1}$, 
$\tilde{\vec r} = \vec r / l$, and the net creation by $k$, $\tilde{r} = r / k$. Second, we refer the bacterial and chemical densities to their initial values, $\tilde{\rho} = \rho / \rho_0$ and $\tilde{c} = c / c_0$, respectively.
Finally, we introduce the rescaled chemotactic length $\tilde{\delta} = \delta /l$ and the rescaled threshold density 
$\tilde{c}_t = c_t / c_0$. This allows us to write the generalized Keller-Segel equations (\ref{eq:dens-KS}) and (\ref{eq:conc-KS}), where chemotactic response is bounded by the hyperbolic tangents, in rescaled form:
\begin{align}
	&\begin{aligned} \label{eq:RescaledDens-KS}
		\frac{\partial \tilde{\rho}}{\partial \tilde{t}} =&\ \tilde{\nabla}^2 \tilde{\rho} + \tilde{r} \tilde{\rho}
		- \chi_0 (1-\langle\cos\beta\rangle) 
		 / \tilde{\delta} \\
       & \qquad \times \tilde{\div}{\left[ \tanh{\left( \frac{\tilde{c}}{\tilde{c}_t}\right)} \tanh{\left(\tilde{\delta}\frac{|\tilde{\grad}{\tilde{c}}|}{\tilde{c}}\right)}
           \tilde{\rho}\ \hat{\vec{s}} \right]}
          \end{aligned}\\[3pt]
\vspace{0.1cm}
	& \begin{aligned} \label{eq:RescaledConc-KS}
		\frac{\partial \tilde{c}}{\partial \tilde{t}} = \frac{D_c}{D_{\textrm{eff}}} \tilde{\nabla}^2 \tilde{c} - \frac{\rho_0}{c_0} \tilde{\rho} \, .
	    \end{aligned}
\end{align}

To arrive at this rescaling, we used $D_{\mathrm{eff}} \approx v_0^2/\omega d$, where we neglected the thermal contribution to 
$D_{\mathrm{eff}}$. The rescaling shows that the generalized Keller-Segel equations are described by a set of six relevant parameters:
$\{ \tfrac{D_c}{D_{\mathrm{eff}}}, \tilde{r}, \chi_0(1-\langle\cos\beta\rangle), \tfrac{\rho_0}{c_0},
\tilde{\delta}, \tilde{c}_t\}$. 


\section{Details of Numerical Solution Scheme}
\label{sec:NumSim}

In the following we study in detail traveling bacterial pulses in an initally uniform density field of a chemoattractant by numerically solving both the polarization extended model (PE)  of Eqs. (\ref{eq:dens}), (\ref{eq:pol}) and (\ref{eq:conc}) and the generalized Keller-Segel model (KS) of Eqs. (\ref{eq:dens-KS}) and (\ref{eq:conc-KS}). The experiments in Ref.\ \cite{saragosti2011directional} are performed in microchannels with cross section $A$. Neglecting any effects at the channel walls, we take all densities to just depend on the $x$ coordinate along the channel. To solve the respective system of equations, we apply a predictor-corrector method at any given time step to efficiently propagate the field variables in time \cite{press2007numerical}. 
As initial field values we choose an exponentially distributed bacterial density, $\rho(x,t=0) = \rho_0\exp(-x/x_0)$, a uniform density of the chemoattractant, $c(x,t = 0)=c_{0}$, and zero polarization $P_x(x,t = 0) = 0$.
During time integration no-flux boundary conditions are employed at $x=0$ and a sufficiently large $x_{\infty}$, $\partial_x \rho\rvert_{0,x_\infty} = 0$ and $\partial_x c\rvert_{0,x_\infty} = 0$, while polarization stays zero, $P_x \rvert_{0,x_\infty} =0$. 
Finally, when integrating Eq.\ (\ref{eq:conc}) the sink term can produce negative concentrations of the chemoattractant\ \cite{rosen1974propagation,rosen1975stability}. To avoid this, we set the 
concentration $c$ to zero whenever it would become negative. This allows the bacteria to fully degrade the chemoattractant without producing negative values for $c$.

\begin{table}[]
\centering
\caption{List of parameters used for the reference pulse in Fig. \ref{fig:propagation} and for the match to the experimental system shown in Fig.\ \ref{fig:Saragosti}.}
\begin{tabular}{llll}
\toprule 
Parameter   &  Value Fig. \ref{fig:propagation} & Value Fig. \ref{fig:Saragosti} & Ref.
\\  \colrule 
  $D$  & $\SI{0.2}{\micro m^2 s^{-1}} $ & same & \cite{berg2008coli} \\
  
  $\Drot$ & $\SI{0.06}{s^{-1}}$ & same & \cite{pohl2017inferring}\\
  
  $r$ & $0$ & $\SI{1.67e-4}{s^{-1}}$ & \cite{saragosti2011directional}\\
  
  $k$ & $\SI{3.35e6}{s^{-1}}$ & same & \cite{adler1969chemoreceptors}\\
  
  $\lequ$ & $\SI{3}{s^{-1}}$ & same & \cite{saragosti2011directional}\\
  
  $v_0$ & $\SI{25}{\micro m s^{-1}}$ & same & \cite{saragosti2011directional}\\
  
  $\langle \cos\beta\rangle$ & $0.392$ & same & \cite{saragosti2011directional} \\
  
  $D_{\mathrm{c}}$ & $\SI{8e2}{\micro m^2s^{-1}}$ & same & \cite{saragosti2011directional}\\
  
  $\chi_0$ & $0.64\ {\lequ\delta}{v_0}^{-1}$ & same & \cite{saragosti2011directional}\\
  
  $\delta$ & $\SI{600}{\micro m}$ & same & \cite{saragosti2011directional}\\
  
  $c_0$ & $\SI{1.26e6}{\micro m^{-3}}$ &  $\SI{2.61e6}{\micro m^{-3}}$  & --- \\
  $c_{t}$ & $10^{-12} c_0$ & $10^{-1} c_0$ & --- \\
  $A$ & $\SI{5e4}{\micro m\square}$ & same & \cite{saragosti2011directional}\\
  
  $x_0$ & $\SI{50}{\micro m}$ & same & \cite{saragosti2011directional}\\ 
  
  $N_0$ & $\SI{1.5e5}{}$ &  $\SI{1.5e5}{}$&  \cite{saragosti2011directional}
  \\ 

\botrule
\end{tabular}
\label{tab:params}
\end{table} 

When we solve our equations with real parameters, we rely on Ref.\ \cite{saragosti2011directional} and take the channel length $x_{\infty} = 10^5 \mathrm{\mu m}$ and the initial decay length of the bacterial density as $x_0= \SI{50}{\micro m}$. This ensures that at $t=0$ $99\%$  of the bacteria can be found within $\SI{200}{\micro m}$ at the channel end at $x_0=0$. We also assume a channel cross section $A = \SI{500}{\micro m} \times \SI{100 }{\micro m}$ to calculate the initial number of bacteria $ N_0= \rho_0 x_0 A$, which we use as a parameter instead of $\rho_0$ in the following. We divide the channel length into $\SI{5e4}{}$ grid points so that the grid length is $\SI{2}{\micro m}$ and use the time step $\SI{0.01}{s}$ for integrating our equations in time. All the relevant parameters are given in Table \ref{tab:params}.
Finally, we will also numerically solve the rescaled Keller-Segel equations\ (\ref{eq:RescaledDens-KS}) and 
(\ref{eq:RescaledConc-KS}) in order to explore the dependence on some of the relevant dimensionless parameters.


\section{Traveling concentration pulses of bacteria}
\label{sec:ChapterIV}

We first introduce the traveling bacterial pulse for a reference system using two values for the initial
number of bacteria, then perform a systematic parameter study, and finally demonstrate  a perfect match with
the experimentally observed bacterial pulse reported in Ref.\ \cite{saragosti2011directional}.


\subsection{Reference system}

\begin{figure}
\includegraphics[width=0.49\textwidth]{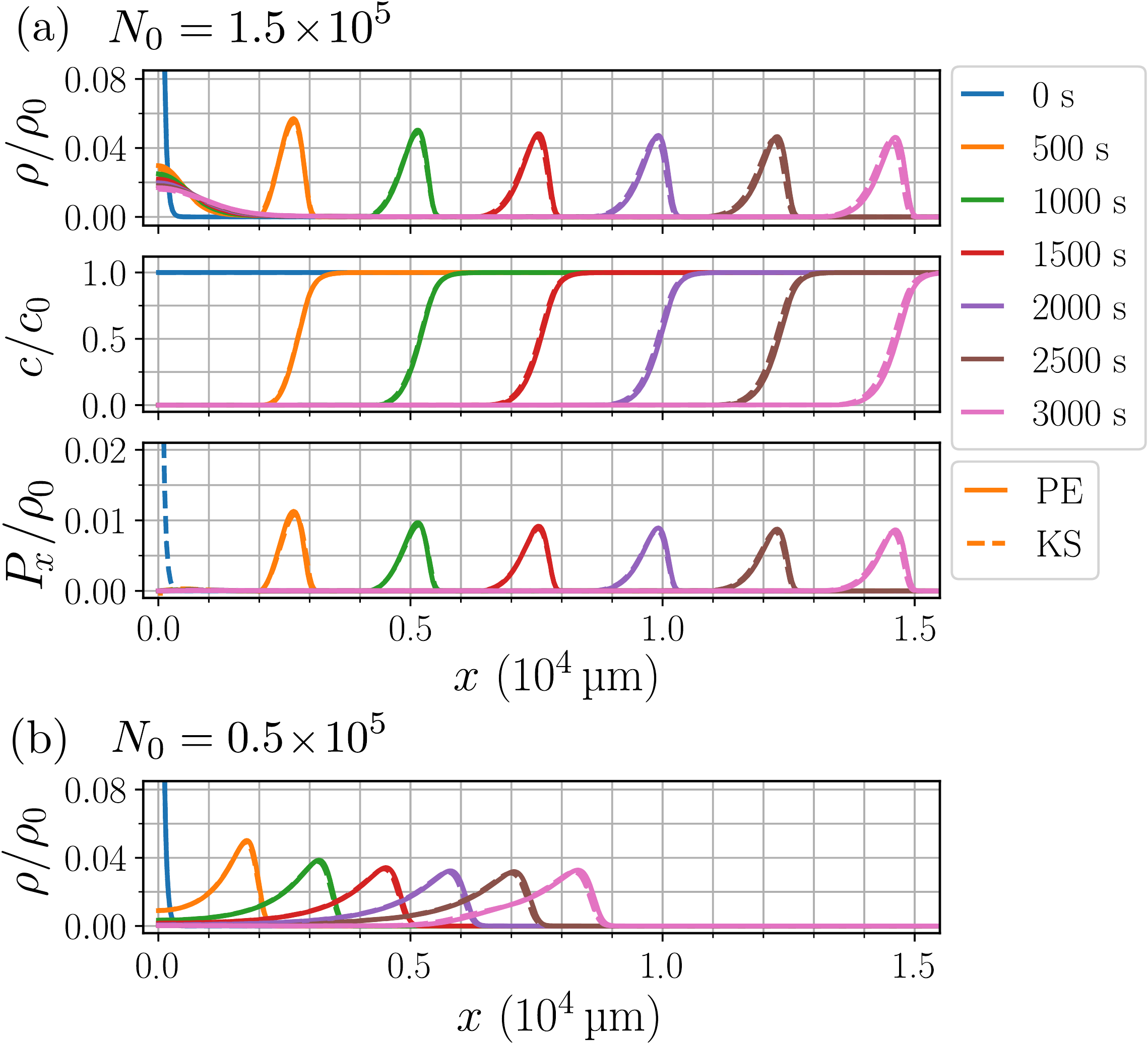}
\caption{Bacterial pulse propagation: (a) Snapshots of the bacterial density $\rho(x)$ (upper panel), the chemoattractant concentration field $c(x)$ (middle panel), and polarization $p_x(x)$ (bottom panel) at equally spaced times for the parameter set given in Table \ref{tab:params}. Solid lines represent the polarization extended model and dashed lines its adiabatic approximation, the generalized Keller-Segel model. (b) Bacterial density for a reduced number of bacteria $N_0 = 0.5 \times 10^5$ compared to the reference system.}
\label{fig:propagation}
\end{figure}
Figure\ \ref{fig:propagation}(a) shows a series of
snapshots of the bacterial density profile $\rho(x,t)$, the concentration field $c(x,t)$, and the polarization $P_x(x,t)$
at equally spaced times for realistic parameters of the \ec/ bacterium listed in Table \ref{tab:params}.
Solid lines represent numerical solutions of the polarization-extended model (PE) and dashed lines the adiabatic approximation leading to the generalized Keller-Segel model (KS). Video S1 of the supplemental material shows an animation of the propagating profiles.

Clearly, while eating up the chemoattractant completely, a traveling pulse in the bacterial density forms that propagates with constant pulse speed $v_p = \SI{4.68}{\micro m s^{-1}}$. It has a comparable width to the traveling step in the chemoattractant profile. In contrast to the bacterial solitons derived from the original KS model in Ref. \cite{keller1971traveling}, our bacterial pulse shows a
small dispersion visible from the slight decrease of the pulse height. It is caused by bacteria that cannot follow the pulse at small chemoattractant concentrations since in our model the chemotactic drift velocity $\vec{v}_{\text{ch}}$ of 
eq.\ (\ref{eq:ChemDrift}) has an upper bound. In contrast, in the original KS model $\vec{v}_{\text{ch}}$ diverges at small chemoattractant concentrations \cite{keller1971traveling}, which allows all bacteria to 
stay in the traveling pulse. Thus, we demonstrate when one allows dispersion a singular chemotactic drift velocity is no longer necessary for traveling pulses to occur. Note, already Ref. \cite{scribner1974numerical} used  the  KS model with a bounded chemotactic drift and observed traveling pulses which was ignored in more recent works \cite{brenner1998physical,saragosti2011directional}. Instead, a second chemoattractant was proposed as explained in the introduction.

One realizes that the profiles generated from the KS and PE model are identical except in the beginning. We start with $P_x = 0$ in the PE model whereas in the KS model a non-zero polarization is calculated from eq.\ (\ref{eq:pol-closed}). It is due to the initial gradient in the bacterial density. Thus, we conclude that the adiabatic limit $\frac{\partial P}{\partial t} \approx 0$ as a condition 
for deriving the generalized KS model is fulfilled. Indeed, the decorrelation or decay time $\omega^{-1} =  \SI{0.49}{s}$ is much smaller than the characteristic time for the pulse propagation. For the latter, we approximately find $\SI{200}{s}$ for the pulse to travel its own width and the polarization always assumes its stationary value.
Our finding also means that the kinetic models of Refs.\ \cite{saragosti2011directional,saragosti2012modeling,emako2016traveling}, which work with the full one-particle distribution function $\psi(\vec{r},\vec{e},t)$, are not necessary to describe pulse propagation. They can be reduced to the Keller-Segel equations.

In Fig.\ \ref{fig:propagation}(a) not all bacteria travel with the pulse but some remain at the initial location. This also occurs in the experiments of Ref.\ \cite{adler1966chemotaxis}. However, the remaining bacteria perform chemotaxis in oxygen as a second 
chemoattractant and thereby initiate a secondary pulse. Since we do not incorporate another chemoattractant, the bacterial distribution at the initial location only broadens by diffusion. Finally, we mention previous
numerical work on the KS model that also showed the bacteria left behind
\cite{scribner1974numerical}. 

In their original work Keller and Segel derived an analytic expression for the speed of the bacterial soliton as a function of the number of bacteria in the soliton $N_p$, the consumption rate $k$, the cross section $A$, and the initial chemoattractant concentration $c_0$ \cite{keller1971traveling},
\begin{equation}
\label{eq:KSpulseSpeed}
  v_p^{th} = \frac{N_p k}{A c_0} \, .
\end{equation}
By integrating the bacterial density along the $x$ direction at time $t = \SI{2000}{s}$, we obtain $N_p = A \int \rho dx = \SI{0.88e5}{}$ bacteria in the pulse and for the number of bacteria left behind close to $x=0$, $N_c = \SI{0.57e5}{}$. Note, $N_p$ and $N_c$ do not add up to $N_0$ since there are also bacteria in the trail between the initial location and the pulse. Using $N_p$ and the parameter values of the reference system from Table \ref{tab:params} in expression\ (\ref{eq:KSpulseSpeed}),
we obtain $v_p^{th} = \SI{4.69}{\micro m s^{-1}}$, which is in very good agreement with our numerical value of $v_p = \SI{4.68}{\micro m s^{-1}}$.

In Fig.\ \ref{fig:propagation}(b) we lower the number of bacteria $N_0$ by a factor of three. Now, all bacteria travel in the pulse and none are left behind. This suggests that the traveling pulse can only carry a certain number of bacteria and thus has a maximum carrying capacity.
We will investigate it in more detail in the following parameter studies, where we use the two traveling pulses from Fig. \ref{fig:propagation} as a reference. For the traveling pulse in Fig.\ \ref{fig:propagation}(b) we determine a smaller
pulse speed of $v_p = \SI{2.66}{\micro m s^{-1}}$. It matches very well with the theoretical prediction from Eq.\ (\ref{eq:KSpulseSpeed}) using $N_p = N_0 = \SI{0.5e5}{}$. Video S2 of the supplemental material shows an animation of the traveling profiles.

\begin{figure*}
\includegraphics[width=0.9\textwidth]{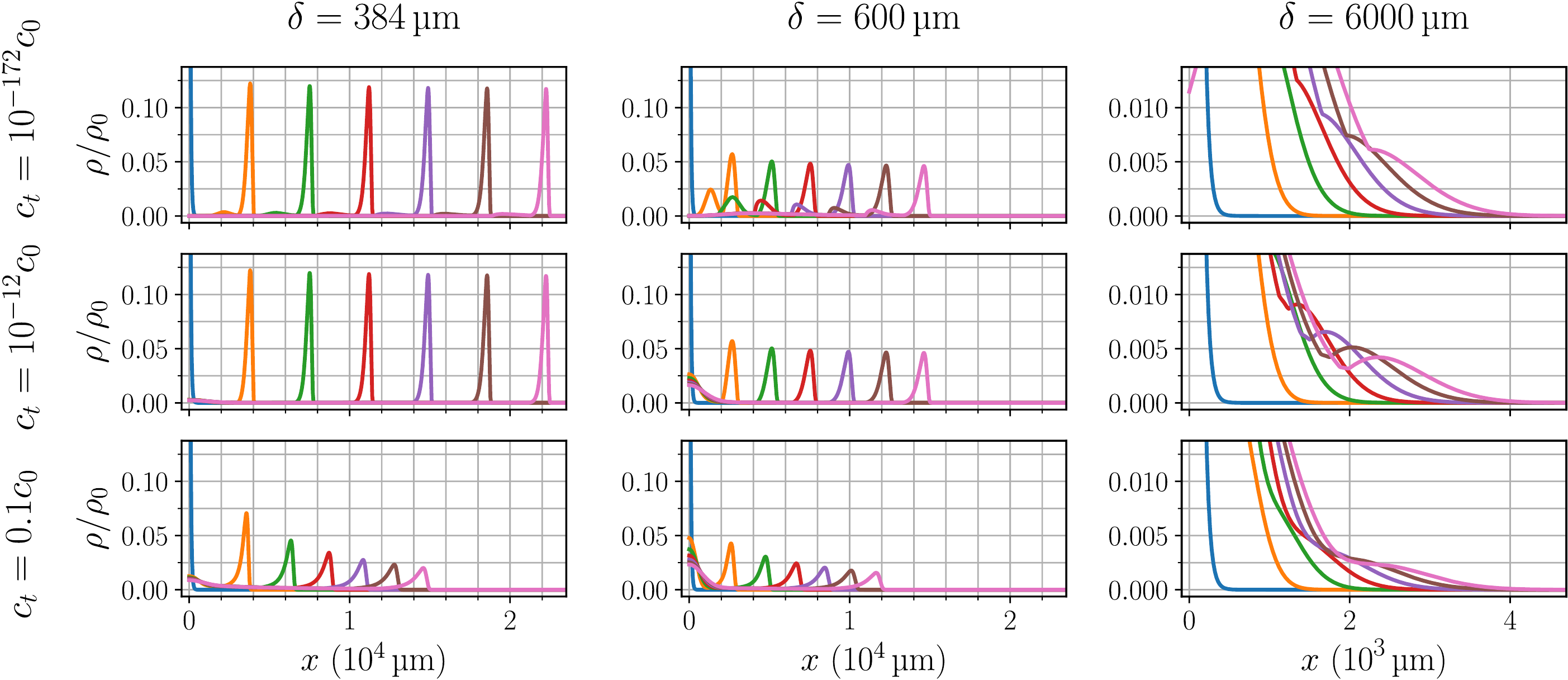}
\caption{Rescaled bacterial density $\rho/\rho_0$ as a function of the chemotactic length $\delta$ and threshold concentration $c_t$. We increase $\delta$ from left to the right and $c_t$ from top to bottom. The reference pulse from Fig \ref{fig:propagation}(a) is in the center. The color code for the different times is the same as in Fig \ref{fig:propagation}. Note the smaller ranges of the $x$ and $y$ axis in the last column.}
\label{fig:ctDeltaStudy}
\end{figure*}

\subsection{Parameter studies}

\subsubsection{Influence of bounded chemotactic drift}
\label{subsubsec.drift}

To keep the tumble rate positive, we introduced the chemotactic length $\delta$ in eq.\ (\ref{eq:tumbleratebounded}), which prevents
the chemotactic drift velocity in eq.\ (\ref{eq:ChemDrift}) to become arbitrarily large. Furthermore, for the chemotactic response the lower 
chemical threshold $c_t$ was  introduced. In Fig.\ \ref{fig:ctDeltaStudy} we explore the influence of both parameters on the traveling bacterial pulse. The chemotactic length $\delta$ increases from left to right and the threshold concentration $c_t$ from top to bottom.
The reference system of Fig.\ \ref{fig:propagation}(a) is in the center. 

A smaller chemotactic length $\delta$ means that the bacterium can sense larger chemical gradients and that the drift velocity 
$\vec{v}_{\text{ch}}$ saturates at a larger value proportional to $\delta^{-1}$. However, it cannot become arbitrarily small since then
the tumble rate in eq.\ (\ref{eq:tumbleratebounded}) becomes negative. In this case our numerical solution scheme is unstable and the bacterial density becomes negative.

The length $\delta = \SI{384}{\micro m}$ is close to the minimal value. In particular for small $c_t$ nearly all bacteria travel in the pulse, 
none are left at the origin. As a result, the pulse travels the fastest. Increasing $\delta$ to $\SI{600}{\micro m}$, 
bacteria left behind are clearly visible. The pulse contains less bacteria and, therefore, is slower. This is also in agreement with the smaller chemotactic drift velocity. Interestingly, for the smallest $c_t$ we observe a second propagating pulse strongly decreasing in height. Finally, if we increase $\delta$ by a factor of $10$ to $\SI{6000}{\micro m}$, the majority of bacteria stay close to the initial location while only a smaller number travels in the pulse (note the 10 times smaller range of the vertical axis). Thus,
the pulse speed is small and the pulse has not yet separated from the non-propagating bacteria.
In conclusion, increasing $\delta$ decreases the maximum carrying capacity significantly and makes the pulse slower.

Increasing the threshold concentration $c_t$ from nearly zero to $c_t = 0.1c_0$ has three effects. First, the dispersion of the pulse increases (especially strongly from the second to the third row) which slows down the pulse. Second, the shape of the pulse becomes more asymmetric as bacteria at the rear flank cannot follow the pulse. Third, the number of bacteria left behind at the initial location increases slightly. In the upper middle plot we recognize that the threshold $c_t$ is so low that the remaining bacteria can still travel by 
chemotaxis, although with a stronger dispersion as the first pulse. Finally, even for the vanishing threshold of the upper left plot a slight dispersion is visible. This again suggests that a true propagating soliton,
for which the pulse shape does not vary in time, is not possible as long as the chemotactic drift velocity is bounded.

\subsubsection{Quantitative study of the rescaled Keller-Segel equations} 
\label{sec:QuantParamStudy}

\begin{figure}
\includegraphics[width=0.49\textwidth]{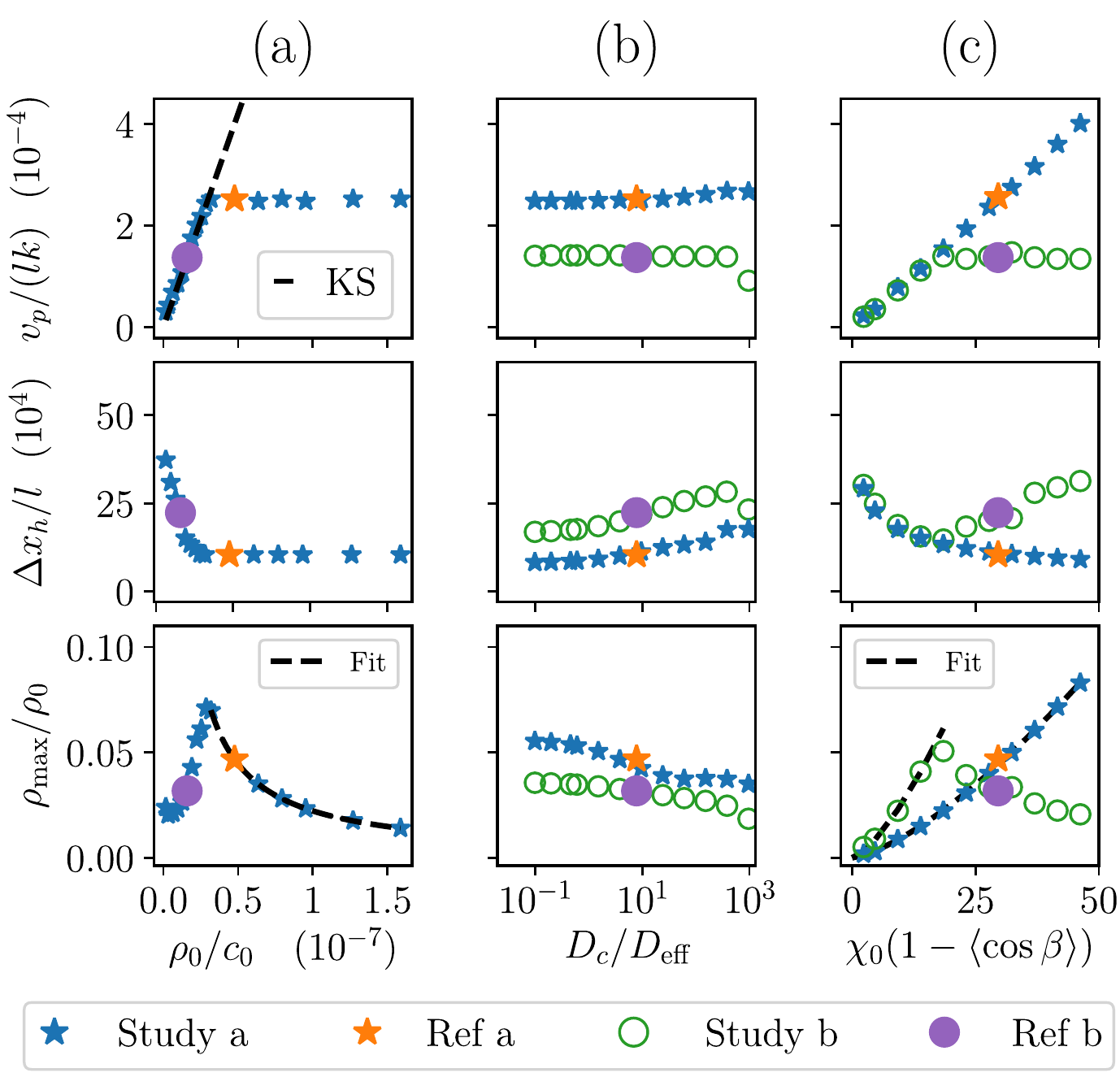}
\caption{Parameter study for the traveling bacterial concentration pulse. The rescaled pulse speed $v_p$ (first line), full width at half maximum of the pulse $\Delta x_h$ (second line), and pulse amplitude $\rho_{\textrm{max}}/\rho_0$ (third line) are plotted as a function of the rescaled parameters $\rho_0/c_0$ (a), $D_c/D_{\textrm{eff}}$ (b), and $\chi_0(1-\langle\cos\beta\rangle)$ (c). 
The values are determined at rescaled time $\tilde{t}= \SI{e10}{}$, where pulse propagation is well established.
When not varied, the following rescaled parameters are used: $\rho_0/c_0 = \SI{4.76e-8}{}$ (blue stars, study a) and $\rho_0/c_0 = \SI{1.59e-8}{}$ (green circles, study b), $D_c/D_{\textrm{eff}} = 7.84$, $\chi_0(1-\langle\cos\beta\rangle) = 29.6$,  $\tilde{\delta} = \SI{1.09e5}{}$, $\tilde{c_t}= \SI{e-12}{}$, and $\tilde{r} = 0$. The reference pulses from 
Fig.\ \ref{fig:propagation}(a) and (b) are marked with orange stars (Ref a) and purple discs (Ref b), respectively. The dashed line in the curve 
$\rho_\mathrm{max}$ versus $\rho_0/c_0$ is a fit to $y(x) = C/x$ with $C= \SI{2.22e-9}{}$ while the dashed lines in the curves 
$\rho_\mathrm{max}$ versus $\chi_0(1-\langle\cos\beta\rangle)$ are fits to $y(x)= Ax^{B}$ with $A = \SI{3.53e-4}{}$, $B=1.42$ (blue stars) 
and $A = \SI{10.4e-4}{}$, $B=1.43$ (green circles).
}
\label{fig:ParamStudyTotal}
\end{figure}

We now consider the rescaled Keller-Segel equations and study the propagating bacterial pulse in detail by plotting the rescaled pulse speed $v_p$, pulse full width at half maximum $\Delta x_h$, and pulse amplitude $\rho_{\textrm{max}}/\rho_0$ as a function of the remaining parameters $\rho_0/c_0$, $D_c/D_{\textrm{eff}}$ and $\chi_0(1-\langle\cos\beta\rangle)$. Again, we neglect bacterial growth by setting
$\tilde{r} = 0$. Figure\ \ref{fig:ParamStudyTotal} shows all results and the relevant parameters are given in the figure caption.

In Fig. \ref{fig:ParamStudyTotal}(a) we see that the pulse speed depends linearly on $\rho_0/c_0$ in agreement with the Keller-Segel 
prediction of Eq.\ (\ref{eq:KSpulseSpeed}) but then saturates at a constant value. The reference pulse from Fig.\ \ref{fig:propagation}(b) 
(purple disc), where no bacteria are left behind at the origin, is located in the linear regime, while the reference pulse from 
Fig.\ \ref{fig:propagation}(a) (orange star), where some bacteria remain close to the origin, propagates in the saturated regime. Thus, in the 
first case adding more bacteria to the system increases the number of bacteria in the traveling pulse and speeds it up. In contrast, in the 
second case additional bacteria remain close to the initial location. Thus, the traveling concentration pulse has a maximum carrying capacity
$N^\ast$ with respect to the amount of bacteria it can hold and all other bacteria are left behind. The transition between both regimes occurs 
at the critical ratio $(\rho_0/c_0)^*$. This allows to discuss the following scenario, for example. Lowering $c_0$  at constant $\rho_0$ speeds
up the pulse in the linear regime since bacteria degregate the chemoattractant faster. However, once $(\rho_0/c_0)^*$ is reached the pulse 
looses bacteria to keep the pulse velocity constant. Thus the carrying capacity of the pulse decreases
for lower $c_0$.

The ratio $\rho_0/c_0$ also influences the pulse shape. In the linear regime with increasing $\rho_0/c_0$ 
the pulse becomes narrower while its absolute height $\rho_\mathrm{max}$ roughly increases with $\rho_0^2$. When reaching the saturated 
regime, the pulse width stays constant as should $\rho_\mathrm{max}$. Thus for the relative height we find 
$\rho_\mathrm{max} / \rho_0 \propto (\rho_0/c_0)^{-1}$. 

In Fig.\ \ref{fig:ParamStudyTotal}(b) we show the pulse speed does not significantly depend on the
ratio of diffusion constants, $D_c/D_{\textrm{eff}}$, for both study cases a (blue stars) and b (green circles). This is in contrast to Ref. \cite{holz1979spatio} where the authors proposed a correction term to Eq.\ (\ref{eq:KSpulseSpeed}), which predicts a decrease of the pulse speed with increasing $D_c$.
However, when examining the bacterial pulse profile, we observe that for larger $D_c/D_{\textrm{eff}}$ the pulse needs longer to form. It needs longer to eat up all the chemoattractant at the origin due to the larger diffusive flux of chemoattractant into the depleted areas. But once the bacteria have fully degraded the chemoattractant, the pulse propagates with the same speed $v_p$ independent of $D_c$. For increasing $D_c/D_{\textrm{eff}}$ the width of the pulse also increases while the amplitude decreases. However, both trends are very small since the quantities do not even change by a factor of two while varying $D_c/D_{\textrm{eff}}$ over four orders of magnitude. Finally, noting the relevant length scale $l= \sqrt{D_{\textrm{eff}}/k}$ to depend on the effective bacterial diffusion constant $D_{\textrm{eff}}$, we find $v_p \propto \sqrt{D_{\textrm{eff}}}$. Moreover, the pulse width increases significantly with $D_{\textrm{eff}}$ while the absolute height increases only slightly.  

Figure \ref{fig:ParamStudyTotal}(c) shows the results for varying the chemotactic parameter $\chi_0(1-\langle\cos\beta\rangle)$. 
For values larger than $50$ the numerical scheme becomes unstable akin to the instability in the chemotactic length $\delta$ already discussed in Sec.\ \ref{subsubsec.drift}.
The pulse speed increases linearly in the chemotactic parameter for the study case a (blue stars) and also for the study case b (green circles) in the range $\chi_0(1-\langle\cos\beta\rangle) < 20$. To understand this finding, we looked in detail at the bacterial profiles. In study case a we find that with increasing $\chi_0(1-\langle\cos\beta\rangle)$ more and more bacteria from the 
vicinity of the initial location enter the pulse, which according to Eq.\  (\ref{eq:KSpulseSpeed}) then speeds up. Thus we conclude for the maximum carrying capacity of the pulse, $N^* \propto\chi_0(1-\langle\cos\beta\rangle$. A similar observation in connection with the scenario of Fig.\ \ref{fig:ParamStudyTotal}(b) gives $N^* \propto \sqrt{D_{\textrm{eff}}}$. Note that our results are in contrast to Ref. \cite{rosen1975analytical} which found $v_p \propto \sqrt{\chi_0}$. In study case b (green circles) we start with a smaller number of bacteria. Thus, at $\chi_0(1-\langle\cos\beta\rangle)  \approx 20$ all bacteria have entered the pulse, which then travels with constant speed for further increase in $\chi_0(1-\langle\cos\beta\rangle)$.
With increasing chemotactic parameter the width of the traveling
pulse decreases in the study case a (blue stars). The curve of study case b (green circles) follows this trend until the pulse speed becomes constant and then steadily increases.

For the pulse amplitude 
of the two study cases the behavior is inverted. The curves are well fitted by $y(x)= Ax^{B}$, where the constants $A$ differ by approximately a factor of three, which is the factor by which the density ratios $\rho_0/c_0$ of both cases differ.
It becomes visible since we plot the reduced amplitude $\rho_\mathrm{max} / \rho_0$. The exponents are nearly the same. The amplitude of the study case b (green circles) decreases for $\chi_0(1-\langle\cos\beta\rangle > 20$ and thereby compensates the increasing pulse width as the number of bacteria in the pulse is constant.

\subsubsection{Influence of growth rate $r$}

\begin{figure}
\includegraphics[width=0.48\textwidth]{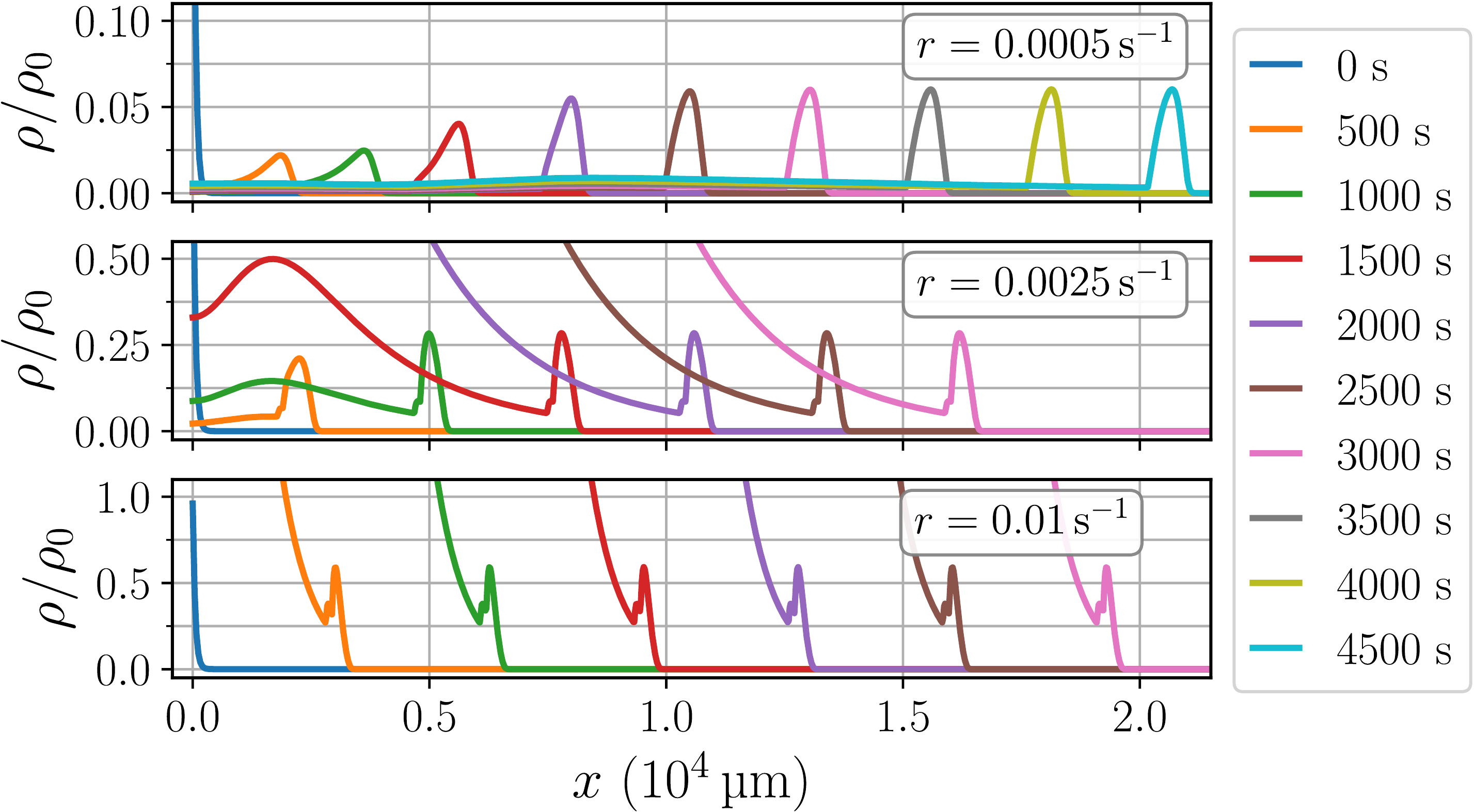}
\caption{Propagating bacterial pulses for three different growth rates $r$. Other parameters are the same as the reference pulse of Fig.\ \ref{fig:propagation}(b) with an initial population of $N_0 = \SI{0.5e5}{}$.}
\label{fig:ParamStudyr}
\end{figure}

Finally, we investigate the influence of the growth term in the Keller-Segel equation\ (\ref{eq:dens-KS}).
Figure\ \ref{fig:ParamStudyr} shows propagating pulses for three different growth rates $r$, 
while the other parameters are chosen as in the reference system but with the reduced initial number of bacteria $N_0 = \SI{0.5e5}{}$. It is below the maximum carrying capacity of the pulse and was used in Fig.\ \ref{fig:propagation}(b). 

Consequently, in the upper panel the pulse grows due to the non-zero growth rate and speeds up in time until the maximum carrying capacity is reached at around $\SI{2600}{s}$. Then, the pulse propagates with 
constant shape like a perfect soliton. However, in our case the pulse leaves a trail of bacteria behind, which originate from the continuous bacterial growth.

In the middle and lower panel, the maximum carrying capacity is reached after $\SI{1000}{s}$ and $\SI{50}{s}$, respectively. Interestingly, the pulse does no longer separate from the broad distribution of bacteria originating from the initial location but rather sits on top of its right flank. In the lower panel the pulse is fastest and its amplitude is highest. This comes from the fact that the broad distribution of the remaining bacteria is much larger compared to the middle panel and more bacteria actively take part in the degregation process of the chemoattractant.  As a consequence, the pulse propagates faster.

Last, we observe that with increasing growth rate the pulse becomes more peaked. This is reminiscent to 
Fig.\ \ref{fig:ParamStudyTotal}(a) and (c), where a faster pulse has a smaller pulse width.

\subsection{Matching the experimental pulse}

\begin{figure}
\includegraphics[width=0.47\textwidth]{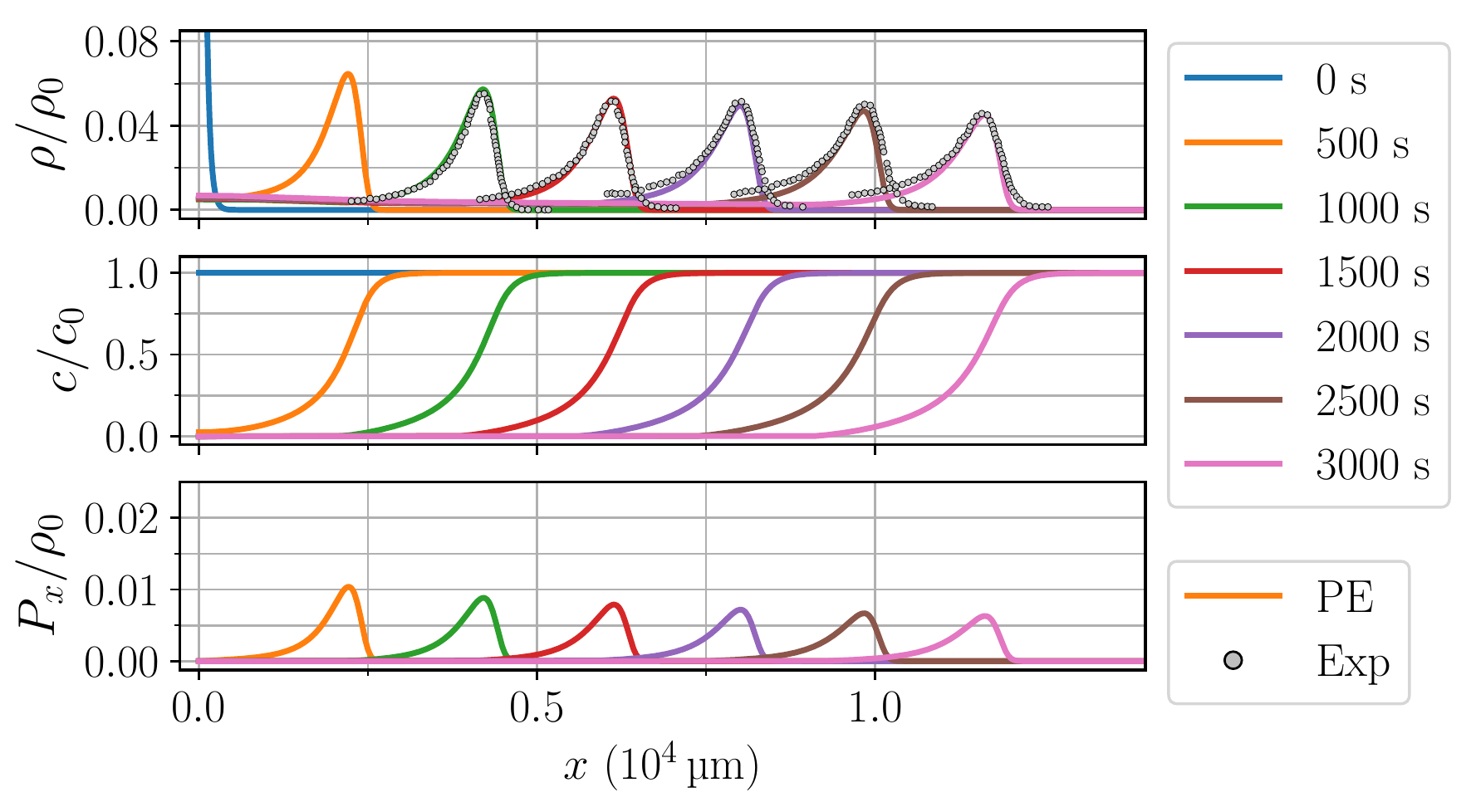}
\caption{Comparison of the experimental traveling pulse (grey circles) from Fig. 1(b) in Ref \cite{saragosti2011directional} and the simulated one using our model (colored lines) with parameters given in Table \ref{tab:params}. Dynamic evolution for different times of both bacterial densities $\rho$ are indicated in the upper panel, the chemical field $c$ in the middle panel and the polarization $P_x$ in the lower panel. The pulse speeds are both $v_p = \SI{3.8}{\micro m s^{-1}}$. 
}
\label{fig:Saragosti}
\end{figure}

Figure \ref{fig:Saragosti} (upper panel) shows the traveling bacterial concentration pulse recorded in the experiments
of Ref. \cite{saragosti2011directional} and compares it to the numerical solution of our polarization extended model.
Both propagating pulses agree very well in shape and in speed $v_p = \SI{3.8}{\micro m s^{-1}}$. We extracted the experimental 
data from Fig. 1(b) in Ref. \cite{saragosti2011directional} and in our model mainly used parameters from the same publication including a non-zero growth rate but also added missing values from Refs.\ \cite{pohl2017inferring} and \cite{adler1969chemoreceptors}.

Moreover, a realistic value for the sensing threshold $c_t = \SI{2.61e5}{\micro m ^{-3}} = 0.1 c_0$ was chosen \cite{adler1969chemoreceptors}. This was necessary to match the asymmetry and dispersion of the pulse. The full parameter set is given in Table \ref{tab:params}.


\section{Discussion and Conclusions}
\label{sec:conclusion}

In the first part of this article we derived the celebrated Keller-Segel equation from first principles using a generalized
Smoluchowski equation for the full distribution function in position and orientation and a multipole expansion. An important
ingredient is the bacterial tumble rate in a chemical field, for which we derived a Markovian response theory from the classical 
chemotaxis strategy based on temporal sensing. Our expression for the tumble rate includes logarithmic sensing, a lower
chemical threshold, and bounds the chemotactic response to keep the tumble rate positive. The multipole expansion provides
a polarization extended model (PE) from which we derived the Keller-Segel equation in the adiabatic limit, where the bacterial
polarization instantly follows variations in the density. We thereby obtain microscopic expressions for the diffusion coefficient
and the chemotactic drift velocity. Due to the bounded chemotactic response the inherent and unrealistic singularity in the drift 
velocity is removed.

Our detailed study of the traveling bacterial pulse shows that its characteristic time is much larger than the relaxation time of 
the bacterial polarization. Thus, PE and KS model provide identical results except for the initial fields and we conclude that
the full Smoluchowski equation as used for example in Refs.\ \cite{saragosti2011directional,emako2016traveling} is not necessary.
This drastically reduces the computational effort and allowed us to perform extensive numerical studies of the bacterial pulse propagation.

We find that due to the upper bound of the chemotactic velocity the traveling pulse can only carry a limited number of
bacteria. To the best of our knowledge such a \emph{maximum carrying capacity} has not yet been reported in the context 
of traveling bacterial pulses. In particular, it is not predicted by the analytic soliton solution of the original Keller-Segel model \cite{keller1971traveling}. Another consequence of the
upper bound of the chemotactic velocity is an effective dispersion of the pulse.  While propagating, 
the pulse leaves a trail of bacteria behind and hence the pulse height decreases and the width expands.
This is consistent with results from Ref. \cite{scribner1974numerical}. The loss of bacteria can be compensated by a non-zero growth rate and we have seen that soliton-like pulses, which propagate with constant shape, are possible.

Exploiting a rescaled version of our KS model, we quantify how pulse speed, pulse width, and pulse amplitude depend on the different unitless parameters. We mention some key results. First, throughout our parameter study we find that the analytic soliton solution of the original KS model still provides a correct estimate for the pulse speed as a function of the number of bacteria in the pulse. Second, we find the maximum carrying capacity to be proportional to the chemotactic strength $\chi_0(1-\langle\cos\beta\rangle)/\tilde{\delta}$ and $\sqrt{D_{\textrm{eff}}}$. As a consequence these parameters affect the pulse speed as long as there are sufficient bacteria in the system so that the maximum carrying capacity is reached.  Third, the diffusion coefficient of the chemoattractant does not influence the pulse speed as predicted in a theoretic model in Ref. \cite{holz1979spatio}. The pulse only takes longer to eat up all the chemoattractant at the origin due to the larger diffusive flux of chemoattractant into depleted areas.

Finally, we show that our simulated pulse propagation is able to match quantitatively the traveling bacterial pulse in the exeriments of Ref.\ \cite{saragosti2011directional} in speed and shape. In contrast to the models used in Refs.\ \cite{brenner1998physical} and \cite{saragosti2011directional}, we do not need a second chemoattractant to generate a traveling concentration pulse as a solution of our generalized KS model.

We mention four directions into which our approach can be extended.
First, so far we did not explore the full PE model. In future works it would be interesting to explore the possibility of having
the bacterial polarization as an independent field and its potential to induce complex dynamics. For example, the alignment 
or polarization of magnetotatic bacteria can be controlled by an external magnetic field \cite{blakemore1975magnetotactic,faivre2008magnetotactic}, which offers the possibility to address polarization as an 
independent field  variable, \emph{e.g.}, by a time-varying external stimulus. The dependence of cell characteristics on 
polarization could also evoke a feedback loop in highly non-linear equations. For example, it was shown that the nutrient 
uptake of bacteria depends on cell shape \cite{young2007bacterial}, meaning that the consumption rate may depend on
the polarization, which then influences chemotaxis \cite{desai2018nutrient}. For Janus colloids with effective phoretic 
repulsion this can generate interesting collective dynamics on times much smaller than the characteristic time scale of the 
bacterial pulse \cite{liebchen2015clustering}. Finally, in complex geometries with characteristic lengths similar to the persistence length of the bacterium, we expect the
polarization equation also to become important.

Second, to describe the multiple pulses that have been observed in experiments with several nutrients 
\cite{adler1966effect,adler1966chemotaxis}, one can extend our model by coupling the bacterial density to several nutrient fields. 
Bacteria that are left behind by the first pulse can then perform chemotaxis in a second nutrient field and thereby create a second pulse.

Third, in our generalized Smoluchowski equation the swimming speed is a constant as it is commonly done for the $\ec/$ bacterium 
also during chemotaxis \cite{berg1972chemotaxis,berg2008coli,seyrich2018statistical}. 
However, some bacteria are known to couple their swimming speed to the concentration of a chemical field \cite{barbara2003bacterial,garren2014bacterial,son2016speed}, a strategy which is called chemokinesis. 
Reference \cite{rein2016collective} derived coupled equations for bacterial polarization and density from a Smoluchowski equation where the swimming velocity depends on the chemical concentration. It is certainly interesting to extend our theory in order to investigate the combined effect of chemokinesis and chemotaxis. A cell that performs both strategies is the sperm cell \cite{eisenbach2004chemotaxis}.

Finally, it would be interesting to extend our approach to chemoattractants to which \ec/ is not perfectly adapted such as serine \cite{berg1972chemotaxis,wong2016role,seyrich2018statistical}. 
For this chemoattractant the mean tumble rate drops as the concentration of serine increases. Thus, when swimming up a chemical gradient, the chemotactic velocity increases \cite{wong2016role}. 
In our generalized KS model, the effective diffusion coefficient of bacteria $D_{\textrm{eff}}$, which directly depends on the tumble rate $\lambda_{\mathrm{equ}}$, now is enhanced in front of the pulse as runs are longer, while it is smaller in the back of the pulse where runs are shorter. This should affect the pulse propagation and indeed, experiments with serine showed that below a certain strength of the chemical gradient traveling pulses do not form \cite{wang1986quasi}.

\section*{ACKNOWLEDGMENTS} Fruitful discussions with J.-T. Kuhr are acknowledged. This work was supported by the German Research Foundation (DFG) within the research training group GRK 1558 and the RISE Germany Program of DAAD.\\


\appendix


\section{Markovian response theory for tumble rate}
\label{sec:MarkovianTumblerate}

Equation (\ref{eq:tumblerate}) describes the non-Markovian linear response of the tumble rate 
on the concentration history of the bacterial trajectory $c(\vec{r}(t'))$. 
Now, by averaging over all possible trajectories that arrive at location $\vec{r}$ with orientation $\vec{e}$, we are able to derive a Markovian response theory for the mean tumble rate. We make use of the fact that rightward and leftward tumbling is equally probable, thus the tumble angle distribution is an even function, $P\left(\vec{e}-\vec{e'}\right) = P\left(\vec{e'} - \vec{e}\right)$. This has indeed been measured for \ec/ in experiments \cite{berg1972chemotaxis,pohl2017inferring,seyrich2018statistical}. In the following, we derive Eq.\ (\ref{eq:tumbleratefinal}) from the main text.

We start with repeating the expression for the tumble rate from the linear response theory: 
\begin{equation}
\label{eq:tumblerateAppendix}
\lambda(t) = \lequ  - \int\limits_{-\infty}^{t}  R(t-t') \, c(\vec{r}(t')) dt'~.
\end{equation}
In the following, we will use three key properties of the response function $R(\tau)$ that were measured in experiments \cite{block1982impulse,segall1986temporal,masson2012noninvasive}.
First, starting from $\tau = 0$  it is non-zero over a time interval $\tau_m \lessapprox \SI{15}{s}$, which we call the memory time. Second, it fulfills $\int_{-\infty}^0 R(\tau)d\tau = 0$, which means the tumble rate does not depend on the absolute chemical concentration (perfect adaptation). Third, it is inversely proportional to the adaptation concentration, $R(t-t') = \widetilde{R}(t-t')/c_a$. Adaption occurs during the memory time, thus we can set $c_a=c(\vec{r}(t))$.
Taking these properties into account, we will use the approximation $\int_{-\infty}^{0}R(\tau)f(\tau)dt \approx 1/c_a \int_{-\tau_m}^0 \tilde{R}(\tau)f(\tau)d\tau$. In particular, due to perfect adaption any additive constant in $f(\tau)$ will not contribute to the integral.

To evaluate Eq.\ (\ref{eq:tumblerateAppendix}), we need an expression for $c(\vec{r}(t'))$. Therefore, we perform a Taylor expansion around the current position $\vec{r}(t)= \vec{r}_t$ and locally approximate the chemical field by
\begin{equation}
\label{eq:approxChemFieldAppendix}
c(\vec{r}') = c(\vec{r}_t) + \grad{c} \cdot(\vec{r}'-\vec{r}_t)~,
\end{equation}
where we used $\vec{r}' = \vec{r}(t')$. In the following derivation, all locations $\vec{r}'$ that contribute to the integral in Eq.\ (\ref{eq:tumblerateAppendix}) should be close to the current location $\vec{r}_t$ so that the linear approximation is valid. Moreover, we assume that temporal variations of the chemical field are negligible within the memory time $\tau_m$ so that  $\grad{c}$ is constant. Both requirements are justified for the bacterial pulse. On the one hand, the mean run length of bacteria
is much smaller than the width of the step in the chemoattractant concentration, and
on the other hand on times comparable to $\tau_m$ the step hardly moves.

Using Eq. (\ref{eq:approxChemFieldAppendix}) in Eq.\ (\ref{eq:approxChemFieldAppendix}), we obtain
\begin{align}
\label{eq:tumblerate2Appendix}
\begin{aligned}
\lambda(t) = &\lequ - \grad{c}/c_a \cdot \int \limits_{t-\tau_m}^{t}  \tilde{R}(t-t') \, \vec{r}(t') dt' ~,
\end{aligned}
\end{align}
where we applied the property of perfect adaption to set 
$\int_{-\tau_m}^{t}  R(t-t') \, \left[c(\vec{r}_t)  - \grad{c}(t)\cdot\vec{r}_t\right]dt' =0$ and that $\grad{c}$ is constant within the memory time $\tau_m$.

To proceed, we write the trajectory of a bacterium that swims with constant velocity $v_0$ along the direction given by unit vector $\vec{e}(t)$ as
\begin{equation}
\label{eq:Langevin1}
    \vec{r}(t') \; = \vec{r}(t) + v_0\int_t^{t'} \vec{e}(t'') dt''~,
\end{equation}
where the bacterium has been at location $\vec{r}(t')$ before reaching $\vec{r}(t)$, thus $t' \le t$. Changes in the swimming
direction due to rotational diffusion are much smaller than due to tumbling. So the trajectory of the bacterium is a sequence
of straight runs and instantaneous tumble events. Denoting tumble events by index $i$, the tumble time $t_i$, and $\vec{e}_i$ 
the direction prior to tumble event $i$, we can write the orientation vector in Eq. (\ref{eq:Langevin1}) at any time $t'' < t$ with the help of a telescope sum:
\begin{equation}
\label{eq:Langevin2}
\vec{e}(t'') =
\vec{e}(t) + \sum_{i=1}^{n} (\vec{e}_i -\vec{e}_{i-1}) = 
 \sum_{i=0}^{n} (\vec{e}_i -\vec{e}_{i-1})
\end{equation}
Here we set $\vec{e}_0 = \vec{e}(t)$ for the current direction after the last tumble event $i=1$ and $\vec{e}_{-1} = \vec{0}$ is used. The number of tumble events in the time interval $t-t''$ is $n$ and we number the tumble events backwards in time.

Now, we determine the mean tumble rate $\langle\lambda(t)\rangle$
by averaging the right-hand side of Eq. (\ref{eq:tumblerate2Appendix}) over an ensemble of bacterial trajectories $\vec{r}(t')$ 
that all reach the position $\vec{r}(t)$ with swimming direction $\vec{e}(t)$. For this, we first have to evaluate  
$\langle\vec{e}(t'')\rangle$ in Eq.\ (\ref{eq:Langevin1}) by averaging over $n$ independent tumble events and considering that $n$ is a random variable. It is determined by the probability distribution $P(n,t-t'')$ of having $n$ tumble events in the time interval $t-t''$.
We can thus write
\begin{equation}
\label{eq:averagedAngleAppendix}
 \langle\vec{e}(t'')\rangle = \sum_{n=0}^{\infty}
 P(n,t-t'')
 \sum_{i=0}^{n} \langle\vec{e}_i -\vec{e}_{i-1}\rangle \, .
\end{equation}
Note, for a constant tumble rate $P(n,t-t'')$ becomes a Poisson distribution. To calculate the mean tumble direction $\langle\vec{e}_i -\vec{e}_{i-1}\rangle$ we use the probability distribution
$P(\vec{e}_{i-1} - \vec{e}_{i})$ from the main text and calculate the first moment as in Eq. (\ref{eq:distavg}) but now with respect to the
incoming direction $\vec{e}_i$ of the tumble event. This gives $\langle\vec{e}_{i} \rangle = \int \vec{e}_{i} P(\vec{e}_{i-1} - \vec{e}_{i})
d\vec{e}_{i} =   \langle\cos\beta\rangle \vec{e}_{i-1}$, where the tumble angle is determined by $\cos\beta = \vec{e}_{i-1} \cdot \vec{e}_{i}$
and we used that $P(\vec{e}_{i-1} - \vec{e}_{i})$ is an even function meaning that left- and rightward tumbles are equally probable.
Repeating the formula for $\langle\vec{e}_{i} \rangle$ for the whole sequence of tumble events, we finally have 
$\langle \vec{e}_i \rangle = \langle \cos\beta\rangle^i\vec{e}(t)$ and the telescope sum in Eq.\ (\ref{eq:averagedAngleAppendix}) 
becomes
\begin{align}
\begin{aligned}
\sum_{i=0}^{n}&\langle\vec{e}_i -\vec{e}_{i-1}\rangle = \sum_{i=0}^{n} \langle \vec{e}_i \rangle - \langle \vec{e}_{i-1} \rangle \\
&=  \vec{e}(t) -\vec{0} + \sum_{i=1}^{n} \langle\cos\beta\rangle^i \vec{e}(t) - \langle\cos\beta\rangle^{i-1} \vec{e}(t) \\
&= \langle\cos\beta\rangle^n \vec{e}(t)
\end{aligned}
\end{align}
Combining the last two equations yields
\begin{equation}
\langle \vec{e}(t'')\rangle = \sum_{n=0}^{\infty}P(n, t-t'') \langle \cos\beta\rangle^n \vec{e}(t)~,
\end{equation}
where the only remaining orientation vector is the current one, $\vec{e}(t)$. 

Now, with the last formula and Eq.\ (\ref{eq:Langevin1}) we can formulate the average location
\begin{align}
\begin{aligned}
 \langle\vec{r}(t')\rangle &= \vec{r}(t) +  v_0\int_t^{t'} \sum_{n=0}^{\infty}P(n, t-t'') \langle \cos\beta\rangle^n dt''~ \vec{e}(t)~.
\end{aligned}
\end{align}
Using it in the tumble rate
(\ref{eq:tumblerate2Appendix}), we finally obtain
\begin{equation}
\label{eq:tumblerateAppendixFinale}
\langle \lambda \rangle = \lequ  - \chi_0 v_0\vec{e}(t)\cdot\frac{\grad{c}}{c_a}
\end{equation}
with
\begin{equation}
\chi_0 =  \int_{t-\tau_m}^t \int_t^{t'} \sum_{n=0}^{\infty}P(n, t-t'') \langle \cos\beta\rangle^n dt'' \tilde{R}(t-t')dt'~.
\end{equation}
Setting $c_a= c(\vec{r}$) we then recover Eq.\ (\ref{eq:tumbleratefinal}) from the main text.


\section{Multipole expansion with bias in the tumble angle}
\label{subsec.app.bias}

Following the same steps as in the multipole expansion without angle bias, we average Eq. (\ref{eq:master-dens}) over all orientations $\vec{e}$ using Eq. (\ref{eq:distnorm-ext}) and obtain
\begin{align} \label{eq:dens-app}
	\frac{\partial\rho}{\partial t} = - \div{ (v_0\vec{P})} \  + D\nabla^2\rho + r\rho \, .
\end{align}
Similarly, we compute the polarization $\int \vec{e} \,$Eq.\ (\ref{eq:master-dens})$\,d\vec{e}$ using Eqs. (\ref{eq:distavg-ext}) and (\ref{eq:distavg-persist}) and with $\hat{\vec{s}}= \frac{\grad{c}}{|\grad{c}|}$ we obtain
\begin{align} \label{eq:polQ-app}
\begin{aligned}
		\frac{\partial\vec{P}}{\partial t} = &-v_0\div{\mymatrix{\mathbf{Q}}} - \frac{v_0}{d}\grad{\rho} + D\nabla^2\vec{P} \\ 
		&-[\Drot(d-1)+(1-\bar{\Theta})\lequ -r]\vec{P} \\ 
		&+(1-\bar{\Theta}_0)\chi\Big(\frac{|\grad{c}|}{c}\Big)\mymatrix{\mathbf{Q}}\cdot\hat{\vec{s}}\\
		&+ \frac{1-\bar{\Theta}_0}{d}\chi\Big(\frac{|\grad{c}|}{c}\Big)\rho\ \hat{\vec{s}}\\
		&+\lequ\sigma\Big(\frac{|\grad{c}|}{c}\Big)\mymatrix{\mathbf{Q}}\cdot\hat{\vec{s}}\\
		&+ \frac{\lequ}{d}\sigma\Big(\frac{|\grad{c}|}{c}\Big)\rho\ \hat{\vec{s}}\\
		&- \chi\Big(\frac{|\grad{c}|}{c}\Big)\sigma\Big(\frac{|\grad{c}|}{c}\Big)s_is_j\int e_ie_j\ \vec{e}\psi(\vec{r},\vec{e},t)d\vec{e}.
\end{aligned}
\end{align}
The last term in Eq. (\ref{eq:polQ-app}) is part of the octupole moment which is defined as
\begin{align}
\mymatrix{O}_{ijk}=\int[e_ie_je_k - \frac{1}{d+2}(e_i\delta_{jk}+e_k\delta_{ij}+e_j\delta_{ki})]\psi(\vec{r},\vec{e},t)d\vec{e},
\end{align}
and represents the interplay between tumble rate variation and tumble angle bias. By neglecting all moments above the first and again defining a relaxation rate $\omega=\Drot(d-1)+(1-\bar{\Theta})\lequ -r$ we arrive at
\begin{align} \label{eq:pol-app}
\begin{aligned}
		\frac{\partial\vec{P}}{\partial t} = 
		& - \omega\vec{P} + D\nabla^2\vec{P} - \frac{v_0}{d}\grad{\rho} \\ 
		& + \frac{1-\bar{\Theta}_0}{d}\chi\Big(\frac{|\grad{c}|}{c}\Big)\rho\frac{\grad{c}}{|\grad{c}|}\\
		& + \frac{\lequ}{d}\sigma\Big(\frac{|\grad{c}|}{c}\Big)\rho\frac{\grad{c}}{|\grad{c}|}\\
		& - \Big[\frac{2}{d+2}\chi\Big(\frac{|\grad{c}|}{c}\Big)\sigma\Big(\frac{|\grad{c}|}{c}\Big)\vec{P}\cdot\frac{\grad{c}}{|\grad{c}|}\Big]\frac{\grad{c}}{|\grad{c}|}\\
		& - \frac{1}{(d+2)}\chi\Big(\frac{|\grad{c}|}{c}\Big)\sigma\Big(\frac{|\grad{c}|}{c}\Big)\vec{P}.
\end{aligned}
\end{align}

\bibliography{bibliography}{}
\end{document}